# Cryptoart: Ethical Challenges of the NFT Revolution[1]

Patrici Calvo
Universitat Jaume I de Castellón
calvop@uji.es

**Abstract.** The digital transformation of the art world has become a revolution for the sector. Cryptoart, based on non-fungible tokens (NFT), is attracting the attention of artists, collectors and enthusiasts for its ability to tokenise any element that can be sold as art in the digital market. That means it is able to become a scarce resource and an economic asset by encapsulating the market value of a piece of digital art, which may or may not have a reference in the real world. This study will delve into the ethical aspects underlying what is known as the NFT Revolution, particularly impacts related to the abuse or destruction of cultural heritage, speculation and the generation of economic bubbles and environmental unsustainability.

### 0. Introduction

The emergence of *blockchain* technology has brought about a revolution for the art world, generating an unprecedented disruptive process whose consequences are being felt strongly in all areas and in all its dimensions: economic, artistic, theoretical, aesthetic, reflexive, sustainable, moral, and so on. In this regard, the emergence of the what is known as cryptoart or cryptographic digital art is particularly outstanding. This is a trend based on the idea of non-fungible tokens (NFTs), smart contracts and cryptocurrencies which, according to its advocates, guarantees the ownership, authenticity, scarcity, exclusiveness, immutability, verifiability and traceability of digital works.

Although there are earlier precedents, cryptoart became established in 2017 in association with the trade in digital works. Until then, the problem with digital works was that they could be quickly be replicated in large quantities on the internet, which meant it was impossible to give them a market value. With the use of blockchain, which covers the ethereum cryptocurrency – tradeable tokens – digital artists managed to endow their works with an aura of scarcity, authenticity and control of ownership, turning them into an NFT: a non-exchangeable resource recognised, encrypted and protected by this cryptographic platform. From that point on, a market value for digital works began to materialise, rising exponentially until 11 March 2021, when Mike Winkelmann, aka Beeple, managed to sell his digital work *Everydays: The First 5000 Days* for €57 million (Fairfield and Trautman, 2021; Joselit, 2021, Dean, 11 March 2021).

[1] This work was supported by Scientific Research and Technological Development Project "Applied Ethics and Reliability for Artificial Intelligence" PID 2019 109078RB-C21 funded by MCIN/AEI/10.13039/ 501100011033, as well as in the development of the European Project "Ethical governance system for RRI in higher education, funding and research centers" [872360] funded by the Horizon 2020 program of the European Commission, as well as the activities of the excellence research group CIPROM/2021/072 of the Valencian Community.

However, the principal players in the art world – artists, academics, gallery owners and so on – have mixed feelings about cryptoart and have expressed a variety of different opinions on it. There are three principal positions: those who believe it is a new bubble; those who think it is a revolution; and those who think the idea has failed.

On one hand, some people closely linked to cryptoart, including cryptoartist Beeple and cryptoinvestor Sundaresan, believe that it is an economic bubble with a clear negative prognosis in the short and medium term. When the bubble bursts, those who have failed to dispose of NFTs will have to pay for the excesses of the sector over the past few years (Ossinger, 6 April, 2021; Collado, 7 June, 2021).

On the other hand, others involved in cryptoart believe that, although it is true that the spectre of an economic bubble haunts the market in NFTs, cryptoart represents a real revolution and an opportunity for the art world far beyond digital art products. It is therefore necessary to adapt as quickly as possible so as not to be excluded from the new artistic context taking shape, which will dominate the sector in the not-too-distant future (Hillmann, 2021; Lydiate, 2021; Nadini et al., 2021; Frye, 2021a; Taylor & Sloane, 2021). As Jeffrey Taylor and Kelsey Sloane (2021), the art world is facing a new paradigm where "In the case of NFTs, the art market remains, but the object, and, for that matter, the work of art, is disappearing. (…) Art markets are disengaging from art and art is disengaging from objects" (Taylor and Sloane, 2021: 170). The influence of NFTs is so great that there are already those who predict the death of art at their hands (Frye, 2021a).

Finally, others are highly critical of the cryptoart movement and the real interests behind it all. They suggest that cryptoart is nothing more than a failed attempt to recreate an artificial perception of scarcity that allows market values to be placed on digital products – a quality that is meaningless in a realm like the digital one where the speed of replication is so high and the copies so similar that it is impossible to establish uniqueness and exclusivity for a digital work. As Brian L. Frye argues, "The problem with 'owning' a 'unique' copy of a digital artwork is there's nothing to own. Obviously, you can own the copyright in a work of authorship fixed in a digital medium. But there is no such thing as a 'unique' copy of a digital file. The concept doesn't even make any sense" (2021a: 4). David Joselit expresses himself in the same critical manner. He believes that the sale of *Everydays*: *The First 5000 Days*: "(…) is particularly perverse given that this 'non-fungible' token is made from images Beeple posted online beginning in 2007, and so in theory could be harvested and collaged from countless iPhones around the world" (Joselit, 2021: 3).

But, beyond these issues, cryptoart also raises ethical challenges which require reflection, such as the abuse or destruction of cultural heritage, the negative impacts on society derived from speculation, the creation of economic bubbles and the generation of economic bubbles and the environmental unsustainability which underlies the world of NFTs. The aim of this study will be precisely to show the ethical aspects underlying cryptoart and its consequences for cultural heritage and society in general. To this end, a brief contextualisation of the process of digital transformation of the art world will be made, with the latest cases of the application and use of AI. Secondly, I will address the emergence and characteristics of cryptoart, a phenomenon supported by blockchain technology which is shaking the foundations of the art market with an influence leaving its mark on all related fields. Thirdly and finally, the ethical challenges underlying the uncritical acceptance of cryptoart will be addressed by looking at case studies and their consequences.

### 1. Artificial Intelligence: robot artists and virtual twins

The digital transformation has burst on to the scene in the art world, with very significant impacts on the processes of generating relevant information and applicable knowledge and producing, distributing and consuming related products and services. AI, one of the three main pillars of the digitisation process, is increasingly being used by artists to complement, develop or even complete their works. This increasingly widespread hybrid practice has generated curiosity among collectors and art enthusiasts, opening an important market niche. AI is also being developed to produce works of art autonomously, without any supervision or additional work from any expert or artist. Along similar lines, the result of automated artistic creation processes based primarily on deep-learning artificial neural networks has captured the attention of the market, generating an increase in the volume of digital art business.

In producing art, the hybrid application and use of AI, where the artificially intelligent algorithm is used by human artists to complement, develop or finish their work, and automated art, where the algorithm or artificially intelligent machine does the work without the interference or supervision of a human artist, are increasingly widespread in the sector, as can be seen with the constant increase in production of this kind, year after year (du Sautoy, 2019).

As far as the art market is concerned, the results of artistic creation processes based entirely or partially on the application of deep-learning artificial neural networks, either in a hybrid form (production generated by the human-algorithm binomial) or automated (production generated exclusively by an algorithm), have managed to capture the attention and interest of collectors and art enthusiasts alike (du Sautoy, 2019). This has generated an increase in the demand for digital art, opening an important market niche where business volume is growing exponentially year after year.

In this regard, there are a number of case studies related both to the production of artworks based on artificially intelligent algorithms and to their sale in the art market. These include the results of the 2016 Rembrandt project; the award of the Lumen Prize Gold Award to a hybrid work in 2018; the first sale of a work entirely made by an artificial neural network in 2018; the first exhibitions of hybrid and automated art in 2019; the presentation of the world's first robot artist in 2019; and above all, the first attempt to generate the *virtual twin* of a recognised artist in 2020.

In the 1970s, artist Harold Cohen introduced AARON, a computer program with tens of thousands of lines, whose task was to create original works of art autonomously. However, AARON was programmed to make top-down decisions based on randomness. In other words, Cohen put "(...) a random number generator at the heart of the decision-making process. (...) to create a sense of autonomy or agency in the machine" (du Sautoy, 2019: 110).

In 2016, the year Cohen died, a facial recognition AI program created by a team of scientists, engineers and historians managed to use 3D printing technology to create an original painting painstakingly mimicking the style of Rembrandt Van Rijn. After analysing 168,263 facial fragments from 326 works by the great Dutch master, the end result was so extremely analogous to the Dutch painter's hand that it was given the title *The Next Rembrandt*, painted 347 years after his death (du Sautoy, 2019: pp. 118-124; Floridi, 2021).

In 2018, artist Mario Klingemann managed to win the prestigious Lumen Prize Gold Award with his work *The Butcher's Son*, in which he applied an algorithm equipped with an artificial neural network. As the artist explains, the creative process of the work consisted of introducing a *stick figure* randomly chosen from the 150,000 human poses available in a database into a *Generative Adversarial Network* (GAN),[2] which began to create the body based on pornographic internet sources (it needed many nude photographs). The body was subsequently optimised with textures, definitions, details, and so on (Elgammal, 2019a; Hertzmann, 2020).

---

[2] An artificial neural network of the GAN type "(...) has two sub networks, a generator and a discriminator. The discriminator has access to a set of images (training images). The discriminator tries to discriminate between 'real' images (from the training set) and 'fake' images generated by the generator. The generator tries to generate images similar to the training set without seeing these images. The generator starts by generating random images and receives a signal from the discriminator whether the discriminator finds them real or fake" (Elgammal, Liu, Elhoseiny and Mazzone, 2017: 5).

That same year, Christie's became the first auction house to sell a canvas painted by an AI algorithm also using GAN technology. The painting, called "Edmond de Belamy, from La Famille de Belamy" and signed by *min G max D Ex[log(D(x))] + Ez[log(1-D(G(z)))]* (an algorithmic formula), was sold for €380,000 (US$432,500) (Arbiza-Goenaga, 2020; Christie's, 2018; Jones, 26 October 2018; Stephensen, 2019;). The creative process of the work basically consisted of borrowing the same artificial neural network used by Klingemann to create his works (GAN), which is open access. It was applied to a database of 15,000 portraits painted between the 14$^{th}$ and 20$^{th}$ centuries, selecting one of the thousands of images generated by the AI and putting a price on it. Despite having had virtually no involvement in the creative process, those responsible for the work –three French programmers with little or no artistic training who call themselves Obvious– stated that the machine doesn't make the art, we do: " "If the artist is the one that creates the image, then that would be the machine (…) 'If the artist is the one that holds the vision and wants to share the message, then that would be us" (Christie's, 2018).

In 2019, an art centre and a gallery became the first venues to hold exhibitions of AI-created works. On one hand, the HG Chelsea gallery in New York exhibited and sold some of the paintings in the *Faceless Portraits* series for 383,000 dollars, a set of pictures painted by an AI algorithm called AICAN (Mansell, 2021). This algorithm, which uses a Contradictory or Anthogonic Creative Network, was primed with 80,000 photographs representing "the Western artistic canon of the last five centuries" (Elgammal, 2019b). As Ahmed Elgammal, one of those responsible for it, argues, "At Rutgers' Art & AI Lab, we created AICAN, a program that could be thought of as a nearly autonomous artist that has learned existing styles and aesthetics and can generate innovate images of its own" (Elgammal, 2019b). On the other hand, the Barbican Centre in London opened the exhibition *AI, More than Human*, where artists from different fields, such as Mario Klingemann, Massive Attack and Es Devlin, showed the results of including AI techniques and technologies in their creative processes (Barbican, 2016; Li, 2020).

That same year, AI combined with robotics to create Ai-Da, the world's first robot artist. With articulated arms, a camera for its eyes and an artificial neural network that uses the *Human in the loop* technique in the learning process,³ the robot is able to create original paintings that look ultra-realistic (Ambrosio, 2019). Created and developed by robotics company Cornish, "Ai-Da does not copy what it sees using a camera, as most robots do. This robot artist interprets what it sees. It creates a 3D image of its surroundings and tries to capture it on paper, applying its own personal touch" (Romic, 2021; Petridou, 2020). It is also worth noting that Ai-Da has already exhibited and its works have been sold at the exhibition *Futures Without Guarantees* held at Saint John's College, Oxford.[4]

---

³ *Human in the loop* refers to a simulation model that involves people in production processes through data and feedback during the training of the machine learning algorithm.

Finally, in 2020, the StyleGAN2 neural network, also called Ganksy, analysed the works of the world's most famous graffiti artist, Banksy, to create and sell original works that perfectly imitate his style on the art market. As its developers argue, "We trained a StyleGAN2 neural network using the portfolio of a certain street artist to create Ganksy, a twisted visual genius whose work reflects our unstable times." Ganksy's works have been well received on the art market, as 175 of the 256 works offered for sale have been placed in just three months.[5] But the idea of Ganksy's creators – the vole.wtf group – goes much further. As Matt Round, one of its developers, states, they intend to give it a robot body "(...) so that it can spray-paint the entire planet" (Castañón, 14 November 2020). This is, then, the first attempt to generate the *virtual twin* of a world-famous artist: his avatar, who, in the future, after a learning process, will paint in his style and on his behalf.

These and many other case studies outline the disruptive power of AI in the complex field of art. It is a power capable of significantly, quickly and continuously transforming, altering or misrepresenting all its aspects and fields of application thanks to its ability to increase reflexiveness in the processes of artistic creativity; the traceability of ideas in the development of new products and research; the permeability of innovative perspectives in creative processes; accessibility to large databases on art; the possibility of using and applying big data to art both in constructing theoretical and applied knowledge and in creating artistic works; economic sustainability; and the attraction of financial resources, among many other things (Agüera. 2017; Fenstermaker, 4 February 2019; López de Mántaras, 2017).

In this sense, the emergence of the what is known as cryptoart is currently the outstanding feature. This is a new way of understanding and managing the art world based on blockchain technology, which allows the democratisation of the digital art world and the immutability, reliability and traceability of its products –non-fungible tokens (NFT) –, which generate virtual scarcity and fix the ownership of a work, and smart contracts, which allow fast, low-cost, two-way relationships between the seller and buyer of virtual work. This is causing rupture and unprecedented instability in the art world. Among other things, digital and artificial artists are migrating towards cryptoart (Grba, 2022; Hong & He, 2021; Liubchenko, 2022; Nair, 10 March, 2022; Shahriar & Hayaw, 2022).

---

[4] In its first exhibition, *Futures without Guarantees*, Ai-Da managed to sell works for a total value of €1.12 million (Romic, 2021; Petridou, 2020).

[5] The prices for the paintings range between £1 and £256. The first painting was sold for £1, and each new purchase costs £1 more than the previous one. As the vole.wtf website says, "All the revenue will help us make more nonsense" (vole.wtf, 2021).

## 2. Cryptoart: blockchain, NFT and cryptographic artists

Today, blockchain technology has become the biggest disruptor in the art world. Although its main utilisation and use can be placed in the subset of the art market, especially in the processes of buying and selling art products, the disruptive imprint of the blockchain is being strongly felt in the way art, creative processes, works of art and artists are understood.

Blockchain was proposed in 2008 by Satoshi Nakamoto —a pseudonym of person or group (Tapscott & Tapscott, 2016). —. "Bitcoin: A Peer-to-Peer Electronic Cash System" (Nakamoto, 2008) presents blockchain as a platform capable of guaranteeing the circulation, traceability and security of an alternative payment system based on peer-to-peer (P2P) technology that allows transactions to be processed electronically without the need for a central authority or a trustworthy system to guarantee it. As Nakamoto argues, it is a peer-to-peer version of electronic cash that allows "(…) online payments to be sent directly from one party to another without going through a financial institution" (Nakamoto, 2008). In this way, the blockchain "(…) marks transactions by converting them into a continuous proof-of-work chain based on hashes, forming a record that cannot be changed without redoing the proof-of-work. (…) Messages are transmitted on a best-effort basis, and nodes can leave and rejoin the network at will, with the longest proof-of-work chain accepted as proof of what happened while they were away" (Nakamoto, 2008).

The blockchain, therefore, represents a kind of open-source ledger whose information is distributed among a large number of users through peer-to-peer technology in an orderly, integrated, transparent, consensual and confidential way (Tapscott & Tapscott, 2016). This means that:

a) every blockchain user becomes a node in the system (decentralisation);
b) every node in the system contains the same information (transparency);
c) none of the information contained in the nodes can be altered in whole or in part (immutability);
d) any alteration of the available information generates a new record or *hash*[6] (traceability); and
e) the availability of information via the hash allows it to be tracked, scrutinised and verified in real time using AI algorithms (auditability).

This revolutionary idea gave rise to a long list of different kinds of cryptocurrencies, such as *bitcoin*, *ethereum*, *litecoins* or *libra*, which has continued to grow. But, more interestingly, it has also sparked a continuous flow of ideas, projects and all kinds of proposals for new forms of practical applications going beyond purely financial transactions and have become highly disruptive elements that are revolutionising all areas of social and human activity. This is particularly noteworthy in the art world, where cryptoart and its fungible tokens (FT) — ERC721— and non-fungible tokens (nft) — ERC20 Standard— are generating unprecedented instability with unpredictable medium- and long-term consequences.

---

[6] *To hash, a* verb that means "to chop" or "to grind", was devised in 1979 by cryptographer Ralph Merkle (Núñez, 2017: 206-207). Its main function in the blockchain is to check the integrity of the information and confirm that it has not been altered (Núñez 2017: p. 204).

Cryptoart, built on blockchain technology, emerged in 2017 linked to the *ethereum* cryptocurrency network and smart contracts, when NFTs created by cryptoartists began to be sold to collectors via art galleries and auction houses by means of *smart contracts*[7] (Magri, 2019; Truby et al., 2022).

As explained in the study "Mapping the NFT Revolution: Market Trends, Trade Networks and Visual Features" (Nadini et al., 2021: 3), "(...) the first example of NFTs used to represent digital art concerns CryptoKitties, a blockchain game on Ethereum that allows players to purchase, collect, breed and sell virtual cats" (2021: 3). Since then their uniqueness and binary nature,[8] but, above all, the guarantees of ownership, authenticity, scarcity, exclusivity, immutability,[9] verifiability and traceability conferred on them by the blockchain system that embraces and protects them, have produced an exponential increase in the market value of NFTs. NFTs therefore constitute "(...) a unique and non-replicable digital assets recorded as cryptographic tokens on the blockchain" (Troby et al., 2022). In other words, the value of NFTs does not depend on their artistic quality or aesthetic capacity, but on their ability to offer the buyer full rights of ownership (binarity) over a unique (scarcity) and traceable (traceability and verifiability) product that is counterfeit-proof (exclusivity) and tamper-proof (immutability and authenticity).

> While NFTs offer possibilities for certifying ownership of various assets, such as real estate, it is the art market where NFTs have proven to be the most popular to date. Tokenising physical or digital art via NFTs has enabled ownership to be both indisputably verified using blockchain technology and provided a simplified means of buying and selling such art. (…) Such content may not have been traditionally thought of as art, but rather part of the digital commons. As such, traders have found a novel means of capitalising on hitherto free content, by creating ownership and selling it as a unique digital asset. The digital signature of each unique NFT makes it a collectible item, proving ownership of a unique piece of art or music, for example (Troby et al., 2022).

---

[7] *Smart contracts* "(…) are user-defined programs that specify rules governing transactions, and that are enforced by a network of peers (assuming the underlying cryptocurrency is secure). In comparison with traditional financial contracts, smart contracts carry the promise of low legal and transaction costs, and can lower the bar of entry for users (Delmolino et al.).
[8] Binary art or artistic binarism means that it is made up only of products that either are (1) or are not (0). There is no possibility of sharing the works through reproductions. This is a negative sum game that reduces the rules of the art world to computing language: a mere mathematical code composed of zeros and ones (binary). This is the only way to be able to positivise the art world: the subjective load on the value of a work is reduced and its objective value is exponentially increased. Works are now supposedly measurable, comparable and auditable, avoiding irrational nonsense that assigns meaningless (mathematical) value.
[9] Immutability is understood as a key to the authenticity of the NFT.

A good example of all this is the market value achieved by the sale of the Nyan Cat *meme* created by the cryptoartist Chris Torres, a drawing of dubious artistic quality and considered to be poor in aesthetic terms, which, despite having already been shared and reproduced millions of times on the internet on websites and social networks, was sold for €500,000 after its conversion and launch as an NFT. But, the really outstanding case involves the cryptographic token *The First 5000 Days*, a collage of 5,000 digital creations put together by the cryptoartist Mike Winkelmann, aka Beeple, which was sold to cryptoinvestor Vignesh Sundaresan via Christie's auction house for €58.5 million Fairfield & Trautman, 2021; Joselit, 2021, Dean, 2021).

> Art projects like Beeple's do not generate this kind of buzz simply because of the cost of the original works, but because of the revolutionary change and impact on the art market and beyond. The fact that owning a work of art requires large sums of money was already a fact – a custom. They are unique objects that wealthy buyers make their own (Dukemon, 2021: 69).

So, the growing trade in cryptoart or cryptographic digital art has been becoming an important subset of the cryptoeconomy, which is itself a subset of what is known as the digital economy. In less than four years, between 23 June 2017 and 27 April 2021, the cryptocurrency managed to generate 6.1 million commercial transactions linked to the purchase and sale of 4.7 million NFTs, with a turnover of €787 million that keeps on growing exponentially (Ante, 2021; Nadini et al., 2021). But, after the sale of Beeple's digital artwork, trade linked to NFTs grew exponentially to reach 40,194,311 million euros per year (fiscal year 2021) in cryptocurrencies linked to ERC-721 and ERC-1155, the two types of Ethereum smart contracts associated with NFT markets and collections (Chainalysis, 2022).

All these issues mean that cryptoart and digital art in general are beginning to be linked both with the democratisation of art and with the emergence of new artistic movements (Cuesta-Valera et al. 2021). The result would be that, on the one hand, the new digital technology is managing to eliminate the barriers that prevent optimal recruitment of talent and, therefore, limit its proper development. Cryptoart allows talent to express itself widely and freely thanks to the accessibility, decentralisation and low cost of the digital market. On the other hand, it also means that cryptoart works are demanding their own, differentiated place among the different art movements. Cryptoart is not only a revolutionary new way of trading art products. It is, above all, a new form of artistic expression that goes beyond the limits of the purely digital sphere, with its own vision, characteristics and style gradually gaining recognition in the art world and imposing a trend (Cetinic and She, 2021). As the cryptoartist Nuno Arteiro says, "NFT art can be considered a new artistic movement making a very strong entry to be recorded in the History of Art", although he does not establish any of its features beyond the technological ones (Arteiro, 3 July 2021).

However, cryptoart also generates problems requiring practical reflection on the negative impacts it generates, or could generate, and the possible procedures for resolving the underlying conflict in order to solve them (Cuesta-Valera et al. 2021). These include issues related to the destruction of artistic heritage, the negative impacts on society derived from speculation and the creation of economic bubbles, the digital divide, and the environmental unsustainability of NFTs, among many other things[10].

### 3. Ethical challenges of cryptoart

There are currently mixed opinions about the future of cryptoart and its short- and medium-term impact on the artistic, economic and social world. Its consequences are not always seen as a revolution in the sector, but rather as an artificial deception for speculative purposes by groups of cryptoinvestors whose negative effects will be visible in the not-too-distant future. In addition, the negative effects resulting from its physical creation, which in some cases go beyond the limits of what is humanly acceptable, are also calling it into question.

#### 3.1 Abuse and destruction of cultural heritage

The success of NFTs, based mainly on the repeated increase in their market value and not on intersubjective agreement on the creative, aesthetic or cultural value of the associated work, has produced trends and behavioural patterns of dubious legal and, above all, moral validity: the abuse or destruction of artistic heritage. The purpose of such violent practices affecting cultural heritage is to increase the market value of a tangible work through its conversion into an NFT and subsequent destruction. In this way, a sense of uniqueness, inviolability and virtual exclusivity is achieved, thus increasing the market value of the work with each new commercial transaction.

In this respect, the most significant case concerns the cryptoinvestors Injective Protocol, who converted graffiti artist Banksy's painting *Morons* into an NFT and subsequently set it on fire live via streaming to "(…) inspire tech enthusiasts and artists". The painting, which before being burned was valued at €80,000, was sold as NFT for €320,000 (García-Madrid, March 20, 2021). In other words, what is being sold now is not an intrinsic quality of the work, but rather its certificate of authenticity.[11] However, this type of behaviour moves away from the intersubjective sphere – the place where, as Jürgen Habermas (1987) would say, human beings come to an agreement about different things about this world through dialogue and the rules of argument – to stray into a dangerous and morally unjustifiable swamp.

---

[10] NFTs not only have an ethical dimension, but also a technical, economic and legal dimension. See other alternative views (Miller, 2009; Marquette de Sousa, 2021) However, this article only addresses the ethical dimension. This is an ethics journal.

In producing art, the hybrid application and use of AI, where the artificially intelligent algorithm is used by human artists to complement, develop or finish their work, and automated art, where the algorithm or artificially intelligent machine does the work without the interference or supervision of a human artist, are increasingly widespread in the sector, as can be seen with the constant increase in production of this kind, year after year (du Sautoy, 2019).

As stated in the 1972 Convention for the Protection of the World Cultural and Natural Heritage, UNESCO considers artistic heritage as cultural property and, therefore, worthy of respect, care and protection by societies and their institutions and organisations (UNESCO, 1972: Art. 1). Similarly, the *Methodological Manual. UNESCO Culture for Development Indicators* (UNESCO, 2014) stresses that goods of artistic interest, such as pictures, paintings and drawings, original works of statuary art and sculpture, original engravings, prints and lithographs, and original artistic assemblages and montages in any material made entirely by hand in any medium or material, are very important for the culture and development of societies, as they constitute "the 'cultural capital' of contemporary societies. It contributes to the continual revalorization of cultures and identities and it is an important vehicle for the transmission of expertise, skills and knowledge between generations. It also provides inspiration for creativity and innovation, which result in contemporary and future cultural products" (UNESCO, 2014: 135).

Therefore, inasmuch as they are universally recognised as an asset, works of art and, in general, all cultural heritage, possess an intrinsic, as well as an extrinsic, value that means they deserve and require respect, care and preservation. In other words, works of art have value regardless of what they make it possible to achieve – economic benefits, user experience, prestige, contemplative pleasure, and so on. So their merely instrumental use erodes the meaning of their existence leading to their decline and disappearance. Therefore, when a work is destroyed in order to increase its extrinsic value – specifically its market price – this flies in the face of the respect it deserves and the care and preservation it requires.

---

[11] The case draws on Bansky her/himself, when he devised a critique of the art market through the live destruction of a reproduction of his work *Girl With Balloon* (Johnston, 6 October, 2018). Although he did not achieve his aim, the work was badly damaged. However, its current value is around £2 million, double what was paid before it was crushed. This prompted many buyers of Bansky's works to think about destroying them to double their market value. According to the art appraisal website MyArtBroker, after the case it received a proposal for the reappraisal of a Banksy lithograph after it had been shredded by its owner (MyArtBroker, 2018). The difference from the *Morons* case, is that the owner did not think of first converting it to an NFT. The destroyed work is one of the 600 lithographic reproductions that make up the collection, now numbering 599. If the owner had previously converted it into an NFT, they would have endowed its reproduction with an authenticity, exclusivity and inviolability the others lack, and, therefore, increased its market value.

3.2 Speculation, fraud and hyperbole

Proponents of cryptoart sell NTFs as elements that are shaking the foundations of the industry to generate something totally new, original and highly beneficial for all stakeholders. As the cryptoinvestor Sundaresan has said, NFTs "(...) herald a new era where technology has allowed artists and collectors around the world to buy and sell art more easily and democratically" (Frank, 30 March 2021). However, many criticisms in this respect focus on the reproduction and even exacerbation of the old problems that maintain the system and limit its development, such as speculation, misappropriation and the creation of bubbles.

On one hand, there are doubts as to whether the cryptoart boom has been due to market demand or whether instead it corresponds to a speculative movement resulting from the connivance of cryptoartists and cryptoinvestors. Sundaresan, for example, has even admitted in an interview that his only reason for acquiring the certificate of authenticity for *Everydays*: *The First 5000 Days* for €57 million was to promote the cryptoartist Beeple. Sundaresan knew better than anyone that his purchase would not change the fact that anyone with internet access can have an exact copy of that work free of charge (Joselit, 2021).

This and other cases have raised the alarm about the high level of speculation supporting cryptoart and its short-, medium- and long-term consequences (Delgado, 10 May, 2021). Despite being seen as a revolution, it does not seem to offer plausible solutions to end the perennial problems suffered by the economy and, specifically, the art market, for decades.

NFTs have managed to fix, certify and trace the ownership and authenticity of digital work, but at the cost of promoting a perverse system, prone to fraud and swindle linked to the usurpation of ideas and the appropriation or illicit exploitation of artistic creations. In this respect, the increase in the number of complaints is significant. One of these came from the actor William Shatner, who denounced the sale of his tweets converted into NFTs. Artist and graphic designer Simon Stålenhag also reported that his GIFs were being converted into NFTs and sold without his permission. The problem with the system being developed by the NFT Revolution is that it is extremely prone to encouraging fraud, which generates controversy and mistrust in the sector (Okonkwo, 2021). This is mainly because the system does not require royalties for an artistic work to be registered. In other words, the system intentionally excludes the existence of intellectual property in order to separate the creation of the work from its digital seal of authenticity. That means what is traded via an NFT – what acquires market value – is not the artistic work in itself, it is its digital record of authenticity. Therefore, what underlies the NFT revolution is not the democratisation of art, but the buying and selling of digital certificates linked to art. Anything is always certifiable, and can be digitally sold, as long as it has not been certified previously.

Finally, linked to the two previous issues, the reasonable doubts about the incredible commercial success and media impact achieved by the NFTs in recent times are worthy of note. Many experts, collectors and artists warn about the economic bubble that underlies the so-called NFT Revolution, comparable to the dotcom bubble that occurred between 1995 and 2000 resulting in a crisis in the industry, with the disappearance of companies, increased layoffs and losses of millions of dollars in the stock market value of the firms affected. As in the past, the incredible increase in the value of NFTs seems to be a response to the strong interference exercised by cryptospeculators on the art market as they seek to profit at the expense of the losses of millions to the crypto-gullible. Various cases occurring in recent years are symptomatic, such as the €2.5 million paid for the NFT of the screenshot of the first 24-character tweet sent by Twitter CEO Jack Dorsey to the social network: "Just setting up my twttr"; the 560,000 dollars paid for the NFT of an open-access column written by Kevin Roose for *The New York Times*; the €612,914 paid for the NFT of the *Nyan Cat* meme; the €57 million paid for the NFT of the Beeple's work *Everydays: The First 5000 Days* and the €450 million cryptoartist Ben Lewis sought to earn by taking a JPG file of the *Salvador Mundi* – a work that some speculators insist on attributing without scientific evidence to Leonardo Da Vinci – modifying it minimally with the image of a left hand holding a fistful of dollars, converting it into an NFT and renaming it *Salvador Metaverdi*; among many other cases (Frye, 2021a). As Joselit argues, "After all, *Everydays* can only be considered non-fungible because enough people agreed that it is. On the contrary, in their previous life, all of this work's constituent images functioned online as a mode of communication" (Joselit, 2021: 4).

Following these astonishing cases, cryptoinvestors such as Sundaresan and specialist journalists such as Roose – shortly after having paid out €57 million for the NFT of Beeple's collage and having sold the *New York Times* column respectively – have warned that NFTs are nothing more than a passing fad that feeds the speculative cryptoart bubble (Kantfish, 2021; White, Wilkoff Yildiz, 2022; Mackenzie & Bērziņa; Collado, 25 March, 2021, 7 April, 2021). And its time may already have come, as Lewis' *Salvador Metaverdi* NFT is currently selling for about $74 on the Open Sea website – a derisory figure considering the cryptoartist's intended price – and has not found a buyer. The problem with speculative bubbles is that they are zero-sum games in which the few who create them usually keep the whole cake and leave the others – especially the most vulnerable in the system – with the obligation to foot the hyperbolic bill for the crisis generated (Calvo, 2018; Case & Deaton, 2020; Miller, 2009).

### 3.3. Environmental unsustainability

The NFT Revolution also conceals a disgraceful side that is intentionally ignored by its most fervent defenders: the serious ecological footprint left by its activity (Truby et al. 2022; Krause & Tolaymat, 2018). Although significant progress has been made in recent years, as David Malone and Karl J. O'Dwyer warned in 2014, when there were only a few cryptocurrency systems and their activity was on a much smaller scale than today, "(...) the energy used by Bitcoin mining is comparable to the Irish national energy consumption" (2014. 4). Today, cryptoart is an environmentally unsustainable system, adding further layers of uncertainty to the already worrying climate change situation and its impact on the planet's natural and social ecosystem (Vries, 2018; Truby et al. 2022). In this respect, the main problem of cryptoart is the high energy costs of NFT tokenisation.

Tokenisation is understood as the process of converting anything into a scarce digital asset through the "(...) encapsulation of value in tradeable units of account, called tokens" (Freni et al., 2020). Thus, through the "(...) creation of a self-governed (tok)economic system, whose rules are programmed by the token issuer" (Freni et al., 2020), tokenisation offers the possibility of creating digital scarcity to it and, at the same time, eliminating the commercial intermediation that makes artworks more expensive, saving time and money in transactions and, therefore, reducing costs and improving profits (Freni et al., 2020, Heredia-Querro, 2021; Krause & Tolaymat, 2018). As stated by Freni et al., "The blockchain made it possible to algorithmically solve the double-spending problem and introduced the concept of digital scarcity, as opposed to the digital abundance characterizing the Internet" (Freni et al., 2020).

However, while tokenisation may avoid traditional transaction costs, it generates new types of transaction costs in the digital world – known as gas costs – that have a very significant environmental impact in the real world. In this respect, it is estimated that currently "Each single Ethereum transaction is estimated to cause 85.47 kgCO2 resulting from the mining devices involved in verifying the transaction, and there were 942,812 NFT sales in the month preceding October 10, 2021" (Truby et al., 2022). In other words, a single NFT can generate many transactions –coining, bidding, transfer of ownership, and so on-. The environmental impact of each of them is equivalent to driving a petrol-driven car for 1,000 kilometres or flying a commercial aeroplane for two hours because of the computational intensity of Ethereum's *Proof of Work* (PoW) algorithm, used to ensure the inviolability and traceability or the work of art (Akten, 2020). As Akten explains, "In less than half a year, the NFTs of various works by an artist have a footprint of 260 MWh: 160 tonnes of $CO_2$ emissions" (Akten, 2020). If the fact that this is data from a year before the 2020 NFT boom is worrying, the lack of interest in finding out its effects is even more so. It is therefore worth reflecting on the need to continue feeding and promoting a system that is so irrationally unsustainable and harmful to the environment and society in general. It may in fact be necessary to look for more environmentally friendly alternatives.

> Furthermore, current emissions from mining devices supporting NFTs transactions are expected to kill people at some time in the future. Bressler estimates that the average lifetime carbon emissions of 3.5 Americans (4434 metric tons or 4,434,000 kgCO2) will kill one person between 2020 and 2100 who would not otherwise have died.28 Death rates from Blockchain transactions can then be estimated based on Bressler's calculations, by calculating the estimated emissions caused by a Blockchain network and dividing the number of transactions to calculate an estimated emissions transaction cost (Truby et al., 2022).

In this respect, it should be noted that the *Proof of Work* protocol used by blockchain projects such as *ethereum*, bitcoin, litecoin and dogecoin, is not the only way to ensure that the system is robust, inviolable and traceable. Other major blockchain projects, such as binance smart chain, cardano, tron and solana[12] use much more environmentally friendly consensus proof-of-stake protocols (David, et al., 2018; Kiayias, 2017; Li, Wei & He, 2020; Tas et al., 2022; Yacovenko, 14 August 2019). Although this is not the ultimate solution, as its ecological footprint remains excessively high and socially and morally intolerable, Proof of Stake (PoS), a consensus algorithm that mitigates the limitations of PoW "(…) by replacing mining action with forging" (Choobineh et al., 2022)[13] stands out among them. As argued by Deuber et al. (2020),

> To mitigate the problems mentioned above, the community investigated alternative consensus mechanisms, based on more energy-efficient resources. One such consensus mechanism is Proof of Stake (PoS) which relies on the rationality of a stakeholder in the system to behave honestly due to the risk of devaluing the currency. In PoS, the consensus leader is chosen solely based on a function of her stake[1] in the system"(Deuber et al., 2020).

There is no doubt that the environmental impact caused by different blockchain projects is one of the major concerns of its developers and those applying it (Choobineh et al., 2022). This problem is, in particular, seen as one of the major constraints to its success and practical application on a large scale. A clear symptom of all this has been the recent announcement by *ethereum*, which, despite its enormous success in recent years, has committed itself to deactivate Proof of Work and move to Proof of Stake in February 2022 (Mourya, 18 October 2021; Fairley 2019). As *Ethereum* has announced on its corporate website, "As part of that roadmap, the existing proof-of-work chain (Eth1) would eventually be deprecated via the difficulty bomb. Users & applications would migrate to a new, proof-of-stake Ethereum chain, known as Eth2" (Ethereum.org, 24 January 2022). New additional protocols, such as Proof of History, and alternative protocols, such as Proof of Hash Ownership, Two-Round PoW and Jakarta PoW, have also been designed and tested in recent years to minimise the environmental impact of blockchain and allow its maximum deployment and potential in practice (Choobineh et al., 2022; Chatbi, 2021a, b, c; Truby et al., 2022).

---

[12] For a list of the different crypto projects and their market capitalisation, see CoinMarketCap (2022).
[13] As argued by Choobineh et al. (2022), "This algorithm ensures lower energy consumption and processing time, although the price paid for this is less decentralisation of the network".

> Nevertheless, a deliberately high energy-intensive proof-of-work blockchain remains the most popular choice for blockchain consensus protocols. Where social pressure fails to persuade developers to switch to more sustainable blockchain, there are a range of available options for policy-makers that can be considered (Truby et al., 2022)

Considering the existing alternatives and the realisation that it is feasible to design and implement increasingly robust and environmentally friendly protocols, designers, implementers and users of blockchain systems are expected to guide their behaviour and decision-making rationally by taking into account values such as sustainability, responsibility and respect for the environment. This is one of the keys to building trust in blockchain and, consequently, to its practical, large-scale application.

**Conclusions**

Ultimately, cryptoart has become a mass phenomenon. It has managed to capture the attention of investors, collectors, consumers and artists in the art world, as well as speculators. And the trend continues to grow despite the latest news pointing to a considerable fall in the market value of NFTs and the possible speculative bubble underlying it.

However, beyond the economic or artistic value of the pieces represented by the NFTs, it is, above all, necessary to question the ethical aspects underlying this revolution in the art world and its negative consequences for society, particularly for the most vulnerable. Among others, this study has focused on three of the main impacts of the NFT Revolution and its consequences: the destruction of artistic heritage, exacerbated speculation and environmental unsustainability. However, there are many fields associated with the digitisation of the art world still to be studied, such as the digital divide, human obsolescence and the opacity of its economic, social and environmental impacts.

Such questions point to the need to reflect on cryptoart's horizon of meaning, the virtues that need to be worked on and promoted by the sector to generate the excellent character that would meet the expectations demanded by a morally mature society, and the conditions of possibility underlying cryptoart and their link with the legitimate expectations at stake. Following Jürgen Habermas (1987), it is worth bearing in mind that only the acceptance by all those involved with cryptoart of the consequences deriving from its normativity and activity can result in the intersubjective agreement that would justify it and give it practical reasons for existence. The art world must therefore begin to seek the agreement of those affected by the establishment and promotion of the communication mechanisms that allow dialogue and possible agreement concerning the NFT Revolution and its economic, social and environmental impacts.